\documentclass[aps,prl,reprint,groupedaddress]{revtex4-2}
\usepackage{graphicx}
\usepackage[english]{babel}
\usepackage{braket}
\usepackage{float}
\usepackage{amsmath}
\usepackage{xcolor,soul}
\usepackage{hanging}
\newcommand\qubitup{\ket{\uparrow}}
\newcommand\qubitdown{\ket{\downarrow}}
\newcommand\qubitupup{\ket{\uparrow\uparrow}}
\newcommand\qubitupdown{\ket{\uparrow\downarrow}}
\newcommand\qubitdownup{\ket{\downarrow\uparrow}}
\newcommand\qubitdowndown{\ket{\downarrow\downarrow}}
\newcommand\Ca{$^{40}\text{Ca}^+$}

\newcommand\infidelity{$6(3) \times 10^{-4}$~} 
\newcommand\fidelity{$0.9994(3)$~}

\begin{document}

\title{High-Fidelity Bell-State Preparation with $^{40}$Ca$^+$ Optical Qubits}

\author{Craig R. Clark}
\author{Holly N. Tinkey}
\author{Brian C. Sawyer}
\author{Adam M. Meier}
\author{Karl A. Burkhardt}
\author{Christopher M. Seck}
\altaffiliation{Present address: Oak Ridge National Laboratory}
\author{Christopher M. Shappert}
\author{Nicholas D. Guise}
\author{Curtis E. Volin}
\altaffiliation{Present address: Honeywell Quantum Solutions}
\author{Spencer D. Fallek}
\author{Harley T. Hayden}
\author{Wade G. Rellergert}
\author{Kenton R. Brown}
\email[]{Corresponding author. kenton.brown@gtri.gatech.edu}
\affiliation{Georgia Tech Research Institute, Atlanta, Georgia 30332, USA}

\begin{abstract}
Entanglement generation in trapped-ion systems has relied thus far on two distinct but related geometric phase gate techniques: Mølmer-Sørensen and light-shift gates. We recently proposed a variant of the light-shift scheme where the qubit levels are separated by an optical frequency [B. C. Sawyer and K. R. Brown, Phys. Rev. A \textbf{103}, 022427 (2021)]. Here we report an experimental demonstration of this entangling gate using a pair of \Ca~ions in a cryogenic surface-electrode ion trap and a commercial, high-power, 532 nm Nd:YAG laser. Generating a Bell state in 35~$\mu$s, we directly measure an infidelity of \infidelity without subtraction of experimental errors. The 532 nm gate laser wavelength suppresses intrinsic photon scattering error to $\sim 1\times10^{-5}$.
\end{abstract}

\maketitle
Generation of 2-qubit entanglement is a key element of universal quantum computing \cite{divincenzo_two-bit_1995} and is typically the most difficult operation to execute with the necessary high fidelity and short duration. The past two decades have seen marked improvement in the measured fidelities for quantum operations across multiple physical qubit platforms, including atomic \cite{bruzewicz_trapped-ion_2019,henriet_2020, saffman_2019}, molecular \cite{lin_2020}, solid-state \cite{chatterjee_2021, kjaergaard_2020}, and photonic systems \cite{arrazola_2021}. The highest-fidelity 1- and 2-qubit operations are currently performed using laser-cooled atomic ions confined in three-dimensional radiofrequency (rf) Paul traps, where 2-qubit gate fidelities of 0.993 (\Ca, 50~$\mu$s gate duration)~\cite{benhelm_towards_2008}, 0.999 ($^{43}$Ca$^+$, 100~$\mu$s gate duration) \cite{ballance_2016}, and 0.9992 ($^9$Be$^+$, 30~$\mu$s gate duration) \cite{gaebler_2016} have been demonstrated with lasers. Recently a related rf-based operation with a fidelity confidence interval of $[0.9983, 1]$ ($^{25}$Mg$^+$, 740~$\mu$s gate duration) was performed in a surface-electrode ion trap~\cite{srinivas_2021}. In Refs.~\cite{ballance_2016, gaebler_2016,srinivas_2021}, the reported 2-qubit gate fidelities are computed from state tomography measurements by correcting for state preparation error and also, in Ref.~\cite{ballance_2016}, for 1-qubit operation errors. In this Letter, we report the generation of a 2-qubit Bell state \cite{nielsen_2010} in 35.2~$\mu$s with a fidelity of \fidelity as measured via Bell-state tomography without correcting for error sources in postprocessing.

Laser-based entanglement generation in trapped-ion systems has until now relied largely on two distinct but related 2-qubit geometric phase gate techniques: M{\o}lmer-S{\o}rensen (MS) \cite{sorensen_2000} and light-shift (LS) \cite{leibfried_2003} gates. Laser-based entangling gates for qubit levels within the $S_{1/2}$ manifold of atomic ions typically require ultraviolet wavelengths for efficient MS or LS gate operations. Alternatively, optical qubits employing narrow atomic transitions (e.g., $S_{1/2}- D_{5/2}$) allow for visible or infrared laser wavelengths for MS gates~\cite{benhelm_towards_2008, akerman_universal_2015}, but with the trade-off that optical phase stability must be maintained throughout the entangling operation.    

We have recently proposed a variant of the LS gate scheme where the qubit levels are separated by an optical frequency and the gate laser wavelength is far detuned from any atomic resonance~\cite{sawyer_2021}. Some advantages of this optical transition dipole force (OTDF) gate include compatibility with dynamical decoupling pulse sequences \cite{leibfried_2003, hayes_2012}, a broad range of feasible entangling gate laser wavelengths (including visible wavelengths), 2-qubit photon scattering error below $10^{-4}$ in some wavelength regimes, and straightforward extension to cotrapped  disparate  species  group-2  ions. As in other optical-qubit systems, our gate is sensitive to the optical phase of the laser used for 1-qubit operations; however, the compatibility of $\sigma^z \sigma^z$ gates with dynamical decoupling pulses that suppress phase errors mitigates this effect. We report here an experimental demonstration of the OTDF gate using a cotrapped pair of \Ca~in a surface-electrode ion trap.

\begin{figure}
	\includegraphics[scale=0.52]{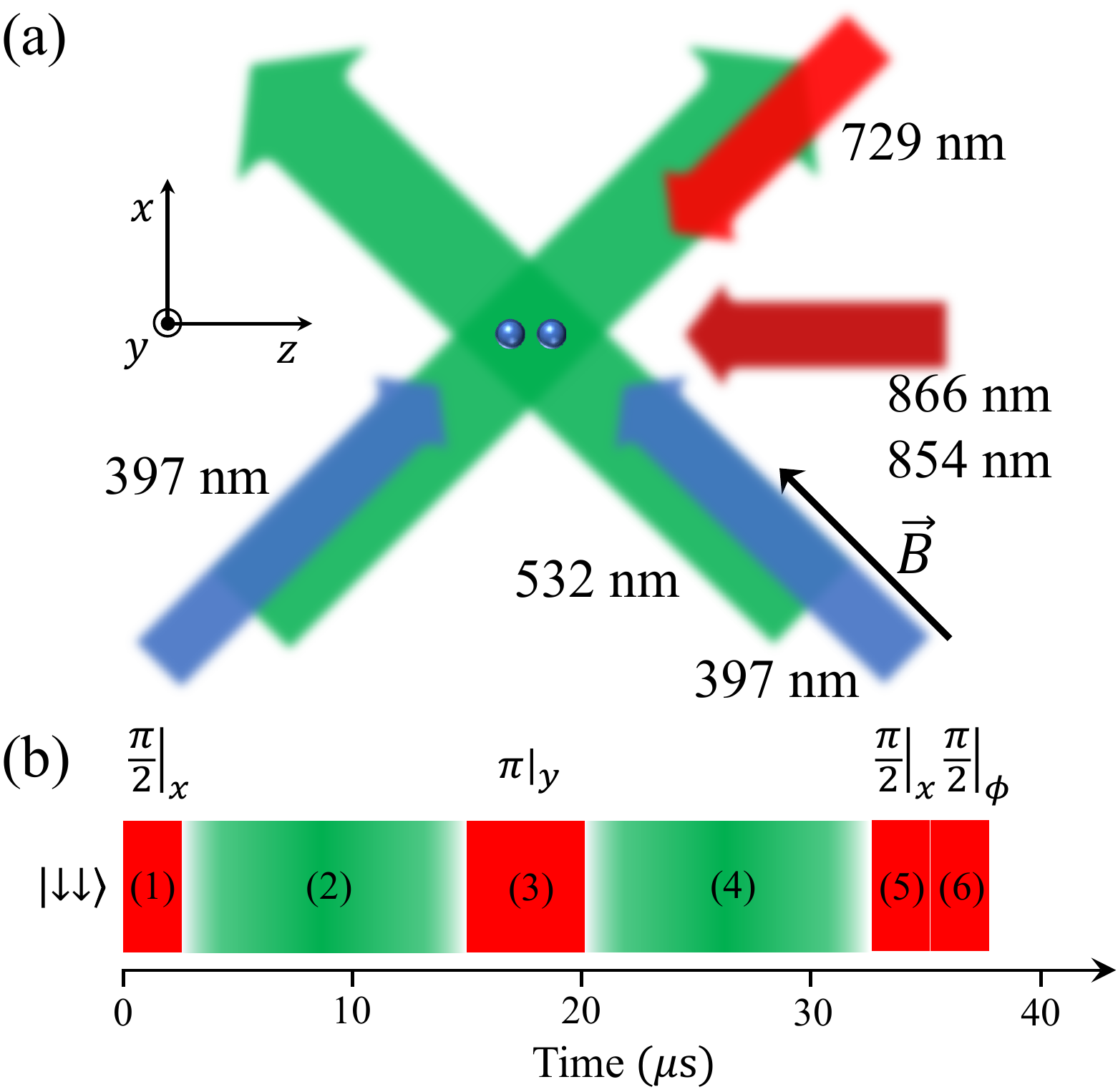}
	\caption{\label{fig:pulse_profiles} (a) Illustration of the experimental orientations of laser beams and the bias magnetic field $\vec{B}$ relative to the two-ion \Ca~crystal. (b) Experimental pulse sequence used in this work for Bell-state generation and parity analysis. Global 1-qubit pulses (1), (3), (5), and (6) are implemented with a resonant 729 nm laser beam. The $\sigma^z \sigma^z$ interactions are produced in pulses (2) and (4) with a pair of 532 nm laser beams intersecting at $90^{\circ}$. The spin-dependent displacements of (2) and (4) are applied with a relative phase offset that minimizes the residual spin-motion entanglement at detuning $\delta$ from the axial breathing mode. The Bell state is created in $35.2~\mu s$ using pulses (1)-(5).}
\end{figure}

The ions are confined 30~$\mu$m above the trap surface in a fixed potential with axial center-of-mass (c.m.) mode frequency $\omega_\mathrm{c.m.}=2\pi \times 2.53~$MHz, breathing-motion (BM) mode frequency $\omega_\mathrm{BM}=2\pi \times 4.38$~MHz, and radial mode frequencies $\sim 2\pi \times 8$~MHz. Confinement transverse to the symmetry axis is achieved via rf potentials applied to the radial electrodes with amplitude $\sim 100$~V at 144~MHz. Static (dc) potentials for confinement along the axis are supplied by digital-to-analog converters and are filtered at the vacuum chamber by 2~Hz low-pass filters to suppress electronic noise that can lead to variations in the trap frequency or in the center of the trap potential~\cite{pino_2020, mccormick_2019}. A magnetic field of 1.07~mT provided via a temperature-stabilized NdFeB permanent magnet outside the vacuum chamber establishes the quantization direction, which is oriented $\sim45^{\circ}$ from the axis as shown in Fig.~\ref{fig:pulse_profiles}.

The ion trap is installed in an ultrahigh-vacuum cryogenic chamber based on a Gifford-McMahon closed-cycle cooler. Cryogenic operation is not a requirement for the gate but is useful to reduce anomalous heating in some cases~\cite{labaziewicz_2008} and to prolong ion lifetimes via cryopumping~\cite{diederich_1998}.
The trap mounting fixture is attached to an optical table below via a series of metal and ceramic standoffs, which provides a solid mechanical reference to the table surface while maintaining sufficient thermal isolation between room temperature, the intermediate stage, and the cold stage. The trap fixture is anchored thermally to the cold stage via copper braids and reaches a temperature of 8.5~K during application of the rf trapping potential. Interferometric measurements reveal trap-mount vibrations along the trap symmetry axis at the level of 30~nm root mean square, with peak excursions up to 80~nm, and dominated by oscillations at the 1.3~Hz cryocooler cycle frequency.

As in Ref.~\cite{sawyer_2021}, we choose $\qubitdown = S_{1/2}~(m_j=1/2)$ and $\qubitup = D_{5/2}~(m_j=3/2)$ as the qubit states.
This optical-qubit transition frequency is linearly sensitive to magnetic field variations, which thereby form a potentially large source of decoherence in our experiments, but the incorporation of a Hahn spin echo into the OTDF gate mitigates this deficiency~\cite{hahn_spin_1950, leibfried_2003}. We perform 1-qubit rotations with a narrow-linewidth 729~nm Ti:sapphire laser locked to an ultrahigh-finesse cavity. The 729~nm laser beam is used both for the spin echo $\pi$ pulse and for the $\pi/2$ pulses of Fig.~\ref{fig:pulse_profiles}(b) employed for Bell-state creation and parity analysis.

We intersect two 532 nm entangling gate beams at an angle of 90$^{\circ}$ [see Fig.~\ref{fig:pulse_profiles}(a)] to create a moving optical lattice that drives the ions with a spin-dependent optical-dipole force (ODF). We choose to generate entanglement with a wavelength of 532~nm, where high-power single-longitudinal-mode lasers are readily available and where photon scattering errors are nearly minimized~\cite{sawyer_2021}.
The wave vector difference between the beams is oriented along the $z$ axis, so that only the motional modes along this direction are driven. Each beam has a waist of approximately 10~$\mu$m and linear polarization along $y$ to maximize optical interference. 

The detuning $(\omega_\mathrm{BM}+\delta)$ between the beams is chosen such that the gate detuning $\delta$ is closer to the BM mode than it is to the c.m. This mitigates the impact of mode heating during the gate (the BM mode heats at $<$1.4~quanta/s compared to 33(14)~quanta/s for the c.m.). For appropriate ODF pulse durations and detunings, it is possible to close the paths of both modes in motional phase space. In practice, $\delta$ differs from the value $2\pi/\tau_\mathrm{ODF}$ that would be chosen for square pulses of duration $\tau_\mathrm{ODF}$, because we employ a ramped intensity profile (3.2~$\mu$s duration) at the beginning and end of the pulses. A power of $\sim100$~mW in each beam achieves the phase gate with two ODF pulses at $\delta\approx 2\pi\times114$~kHz and  $\tau_\mathrm{ODF}=12~\mu$s.

The ODF beams are created from a single seed laser whose intensity is controlled via an acousto-optic modulator (AOM). Using this AOM, we ramp the intensities of both beams in tandem with sine-squared tapers at the beginning and end of each ODF pulse; such adiabatic ramping is important to suppress sensitivity to the absolute phase of the optical lattice~\cite{sawyer_2021, ballance_2016}. Each of the ODF beams passes in turn through an additional AOM, so that its frequency and phase can be controlled independently and its intensity can be stabilized via monitoring and feedback. 

We use the Ramsey sequence depicted in Fig.~\ref{fig:pulse_profiles}(b) to translate the spin-dependent phase shifts imparted from the ODF interaction into population imbalances that can be determined via optical fluorescence detection. First the ions are Doppler laser cooled with beams at 397~nm and 866~nm. Then the BM and c.m. modes are cooled below 0.1 quanta via pulsed resolved-sideband cooling on the $S_{1/2}~(m_j=-1/2)\rightarrow D_{5/2}~(m_j=-5/2)$ transition. The ions are then initialized to $S_{1/2}~(m_j=1/2)$  via a combination of 397 nm excitation with circularly polarized light and seven subsequent iterations of frequency-selective two-step optical pumping through the $D_{5/2}$ and $P_{3/2}$ levels~\cite{suppinfo}. With the qubits now in the state $\qubitdowndown$, the experiment follows the Ramsey sequence diagrammed in Fig.~\ref{fig:pulse_profiles}(b): a first $\pi/2$ pulse (1) creates a superposition of all four 2-qubit states.
The ODF beams are then applied (2) for a duration $\tau_\mathrm{ODF}$ to drive the ions around a closed trajectory in motional phase space, thereby imparting a spin-dependent geometric phase. To symmetrize the phase imparted onto each state within a given parity subspace, the qubits are flipped with a $\pi$ pulse (3) and subsequently driven with a second ODF pulse (4) nominally identical to the first. However, the ODF phase of (4) is adjusted to match the initial phase of (2) in the rotating frame of the qubit. A second $\pi/2$ pulse (5) then terminates the Ramsey sequence to create the desired Bell state. An optional third $\pi/2$ pulse with variable phase (6) is used for tomographic analysis. Finally, fluorescence at 397~nm is collected from both ions simultaneously for a duration of 100 or 200~$\mu$s (see Supplemental Material \cite{suppinfo}), and the experiment is repeated to build histograms of detected photon counts.

The resulting photon-count histograms are well approximated as a weighted sum of three Poissonian histograms (a two-parameter probability mass function) corresponding to two bright ions ($\qubitdowndown$), a single bright ion ($\qubitdownup$ and $\qubitupdown$), and two dark ions ($\qubitupup$). We determine the populations in these three subspaces ($P_0$, $P_1$, and $P_2$, respectively) by maximizing the likelihood of a given observed histogram within this two-parameter model (see Supplemental Material \cite{suppinfo}\nocite{nie_2009}).

To properly suppress errors due to gate detuning fluctuations, it is important that the spacing between the ions be an integer multiple of the lattice wavelength. If this were not the case, the ODF on $\qubitdowndown$ would not match that on $\qubitupup$, and Walsh modulation would not be achieved via the spin echo \cite{hayes_2012,sawyer_2021}.
For this calibration, we apply an on-resonance ODF pulse ($\delta=0$) to ions initialized in $\qubitdowndown$ and look for resulting motion in the BM mode as quantified by excitation of its red motional sideband. We then vary the axial confinement strength and repeat this process so as to minimize the observed excitation.
Similarly, the ODF intensity on the two ions should be matched for optimum performance. This is achieved by performing a Ramsey experiment with only a single ODF beam illuminating the ions and maximizing the beat note period observed in the populations as the length of the Ramsey experiment is varied.

\begin{figure}
	\includegraphics{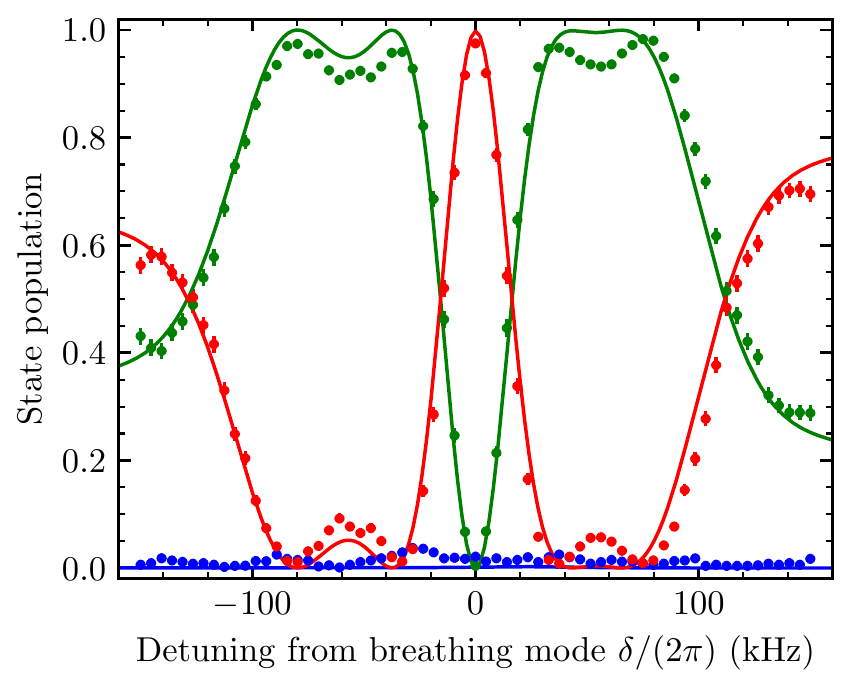}
	\caption{\label{fig:detuning_scan} Measured (points with error bars) and simulated (solid lines) two-ion populations for varying optical-dipole-force detuning $\delta$ from the axial breathing motional mode. We measure the two-ion bright (green, $P_0^\prime$), two-ion dark (red, $P_2^\prime$), and one-ion bright (blue, $P_1^\prime$) populations at each detuning value (error bars represent the 68\% confidence interval assuming binomial statistics). The simulated populations are obtained via numerical integration of the Schr\"{o}dinger equation, including non-Lamb-Dicke effects.}
\end{figure}

For a fixed ODF interaction time, we choose a gate detuning $\delta$ to maximize the entanglement fidelity. Figure~\ref{fig:detuning_scan} shows state populations $P_0^\prime$, $P_1^\prime$, and $P_2^\prime$ (the prime notation denotes populations measured before the final analysis pulse) after the second $\pi/2$ pulse (5) as $\delta$ is varied; here we intentionally compensate for the change in phase between the two ODF pulses for each $\delta$, so the ODF should have the same phase (in the rotating frame) at the start of each pulse, and $P_1^\prime$ is minimized accordingly for all gate detunings.
Experimental data are represented as points with error bars, and the solid lines are theoretical predictions with a 2~kHz offset in detuning as a free parameter. 
The optimum gate detuning lies near $\delta=2\pi\times 114$~kHz, where $P_0^\prime$ and $P_2^\prime$ are measured with equal probability. For gate calibration, we use an analytic estimate of the ideal detuning based on the chosen pulse durations and shapes; then we sweep the intensity of the ODF beams to ensure that the even-parity populations are matched for this choice of $\delta$.

\begin{figure}
	\includegraphics{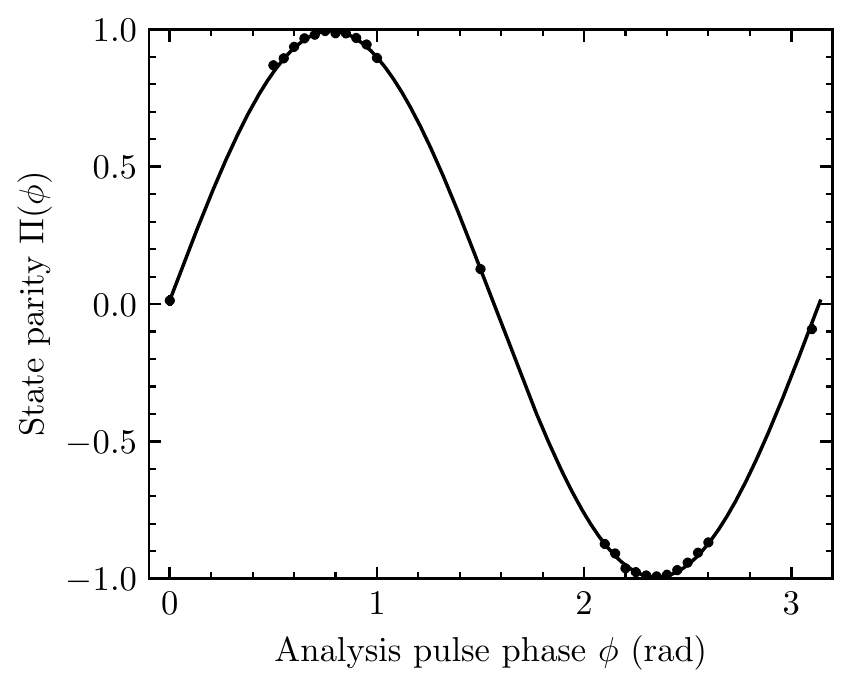}
	\caption{\label{fig:parity_scan} Two-ion parity measurements (points with error bars) and sinusoidal fit (solid line) versus analysis phase (error bars represent the 68\% confidence interval assuming binomial statistics). The density of measured points is highest near the peaks of the parity oscillation. In some cases, the error bar is smaller than the plot marker.}
\end{figure}

\begin{figure}
	\includegraphics{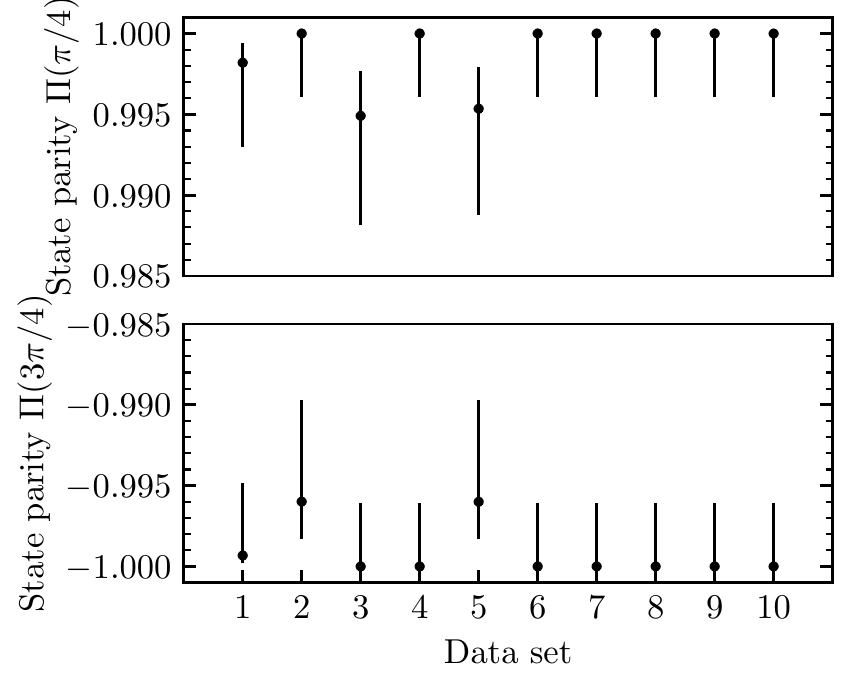}
	\caption{\label{fig:parity_amplitude} Repeated measurements of the two-ion parity for alternating values of analysis phase corresponding to the peak amplitudes of Fig.~\ref{fig:parity_scan} (error bars represent the 68\% confidence interval assuming binomial statistics).}
\end{figure}

Figures~\ref{fig:parity_scan} and~\ref{fig:parity_amplitude} show the results of performing this experiment for varying values of the analysis pulse (6) phase $\phi$. Figure~\ref{fig:parity_scan} gives the parity $\Pi(\phi) = P_0+P_2-P_1=1-2 P_1$ of the observed final state as a function of $\phi$, revealing the expected periodicity of $\pi$ for a two-spin Bell state.
The parity very nearly approaches 1 near $\phi=\pi/4$ and -1 near $\phi=3 \pi/4$. To measure the maximum and minimum values of this curve more efficiently, the experiment was repeated $10\,000$ times, alternating for each repetition between these two discrete phase values. To look for possible time-dependence in the experiments, we binned the results into ten data sets each with $2\times 500$ repetitions and determined the parity of each data set (Fig.~\ref{fig:parity_amplitude}).
Here we use a Jeffrey's interval to estimate the error bars for each point \cite{brown_2001}, because the data sets are too small for resampling to yield a meaningful error bar. Six of the data sets are consistent with unity parity amplitude, while four exhibit a small reduction.

To establish the most precise value of the parity amplitude from these data, we determine the parity of all $5\,000$ repetitions at $\phi=\pi/4$ ($0.999\,02$) and at $\phi=3\pi/4$ ($-0.999\,20$), yielding a parity amplitude of $A=0.999\,11$. To estimate the uncertainty of this value, we perform bootstrap resampling as follows. Each experiment gives a measured number of photon counts. All of the counts for $\phi=\pi/4$ are binned into a single data set with $5\,000$ results, and all of the counts for $\phi=3\pi/4$ are binned into another data set. We then generate $10\,000$ bootstrap data sets by resampling from the experimental data sets with replacement, and we analyze each such bootstrapped data set as described above. The mean of the resulting distribution is $0.999\,10$ with a $68$\% confidence interval of $[0.998\,53,0.999\,61]$, showing that the resampling is not biased significantly.

\begin{figure}
	\includegraphics{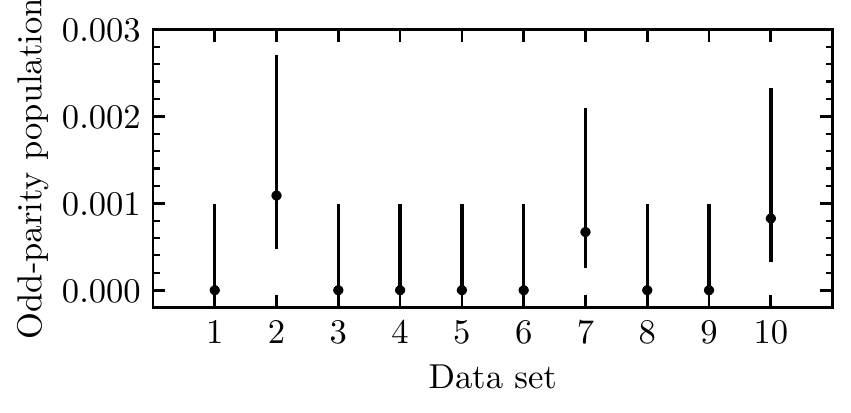}
	\caption{\label{fig:population_scan} Repeated measurements of the one-ion bright population $P_1^\prime$ before a parity analysis pulse (error bars represent the 68\% confidence interval assuming binomial statistics).}
\end{figure}

Similar experiments without an analysis pulse (6) are also performed. Here we repeat the experiment $10\,000$ times, bin the results into ten data sets each with $1\,000$ repetitions, and determine the residual odd-parity population $P_1^\prime$ in each data set (Fig.~\ref{fig:population_scan}).
Seven of the data sets are consistent with $P_1^\prime=0$, while three exhibit a small deviation. Treating all $10\,000$ repetitions as a single data set yields $P_0^\prime+P_2^\prime=1-P_1^\prime=0.999\,77$. Resampling as before from this data set to create $10\,000$ bootstraps gives a distribution with a mean of $0.999\,76$ and a 68\% confidence interval of $[0.999\,56,1.000\,00]$, again showing that the distribution is not biased.

From these measurements we can obtain an estimate of the Bell-state fidelity $F$ via the relation $F=\frac{1}{2}(P_0^\prime+P_2^\prime+A) = 0.999\,44$ \cite{leibfried_2003}. Using the same $10\,000$ resampled data sets that were already generated to estimate the parity amplitude $A$ and the populations $P_0^\prime+P_2^\prime$, we generate a distribution of bootstrapped fidelities (see Supplemental Material \cite{suppinfo}).
The mean of this distribution is $0.999\,43$ with a 68\% confidence interval of $[0.999\,13,0.999\,73]$, corresponding to an infidelity of \infidelity. 

\begin{table}
\centering
\begin{tabular}{||l | c||} 
\hline
Error source & Contribution $(\times10^{-4})$ \\ 
\hline\hline
\parbox[t]{2in}{\raggedright Spin dephasing} & $< 4.1$ \\
Metastable $D_{5/2}$ decay & 0.6 \\
Detection $D_{5/2}$ decay & 0.9 \\
Finite axial mode temperature & 0.2 \\
Spontaneous photon scattering & 0.1 \\ 
BM mode heating & 0.07 \\
Trap frequency variation & 0.01 \\
\hline
\end{tabular}
\caption{Estimated dominant contributions to Bell-state infidelity. The total error due to spin dephasing is bounded by the difference between the measured Bell-state infidelity and the sum of all other known errors.}
\label{table:error_budget}
\end{table}

Estimated dominant contributions to the infidelity are summarized in Table~\ref{table:error_budget}. The first line in the table is an upper bound defined as the difference between our measured Bell-state infidelity and the sum of all separately quantified errors. We report this value here as an upper bound on spin dephasing (i.e., phase and intensity instability of the 729~nm laser beam, intensity instability of the 532~nm laser beams, and fast ambient magnetic field noise). The largest contribution is error from the four 1-qubit rotation pulses (1), (3), (5), and (6). Randomized benchmarking measurements of our 1-qubit operations shortly after the Bell-state experiments reveal an error-per-gate of $1\times 10^{-4}$, so that a naive summation of the errors on our two qubits from these four pulses gives a value of $8\times 10^{-4}$. This is clearly an overestimate of their impact on our entangling gate experiments, but is the same order of magnitude as the upper bound of Table~\ref{table:error_budget}.
A more detailed discussion of the various error sources is found in the Supplemental Material \cite{suppinfo}.

In conclusion, we have demonstrated an OTDF entangling gate using a pair of \Ca~ions in a surface-electrode ion trap, measuring a Bell-state fidelity of \fidelity via parity analysis. Of the estimated experimental error sources, global 1-qubit operations constitute the largest single error by roughly one order of magnitude. By contrast, intrinsic sources of decoherence (spontaneous photon scattering at 532 nm and metastable $D$-state decay) impart a combined error of $\sim7\times10^{-5}$. Future improvements in the frequency stability of the 729 nm laser system toward the state of the art~\cite{Bermudez_assessing_2017-1} should significantly increase our Bell-state fidelity. Straightforward extension of the OTDF technique to multispecies ion crystals is detailed in Ref.~\cite{sawyer_2021}, and implementation of this scheme using radial modes would allow entanglement generation within longer ion chains. Operation at longer (i.e., infrared) wavelengths is also an interesting avenue for further exploration.

This work was done in collaboration with Los Alamos National Laboratory.

%\bibliographystyle{apsrev4-2}
%\bibliography{experimental_gate_paper}

%

\clearpage
\section{Supplemental Material}
\setcounter{page}{1}
\subsection{Determination of State Populations}
Our photon-count histograms are well approximated as a weighted sum of three Poissonians (a two-parameter probability mass function, because the three weights $P_2$, $P_1$, and $P_0$ must add up to unity). These three Poissonians correspond to two bright ions ($\qubitdowndown$), a single bright ion ($\qubitdownup$ and $\qubitupdown$), and two dark ions ($\qubitupup$). If we define $k_2$, $k_1$, and $k_0$ as the mean number of counts in each Poissonian (determined via independent measurements and maintained at fixed values during the following fitting procedure), we can write the probability mass function (PMF) as
\begin{equation*}
\begin{split}
    \mathrm{PMF}(n;P_0,P_1)&=P_0\frac{k_0^n \exp{(-k_0)}}{n!}+P_1\frac{k_1^n \exp{(-k_1)}}{n!} \\
    &+(1-P_1-P_0)\frac{k_2^n \exp{(-k_2)}}{n!}.
\end{split}
\end{equation*}
For a given dataset of identical experiments, with each repetition in the dataset labeled by $i$, we have a set of measured photon counts $\{n_i\}$. With this we define the usual log-likelihood as
\begin{equation*}
\begin{split}
    l(P_0,P_1)=\sum_{i} \ln(\mathrm{PMF}(n_i;P_0,P_1)).
\end{split}
\end{equation*}
We then numerically maximize this function to find the most likely values for $P_0$, $P_1$, and $P_2=1-P_1-P_0$.

\subsection{Fidelity Distribution}
\begin{figure}[htp]
	\includegraphics{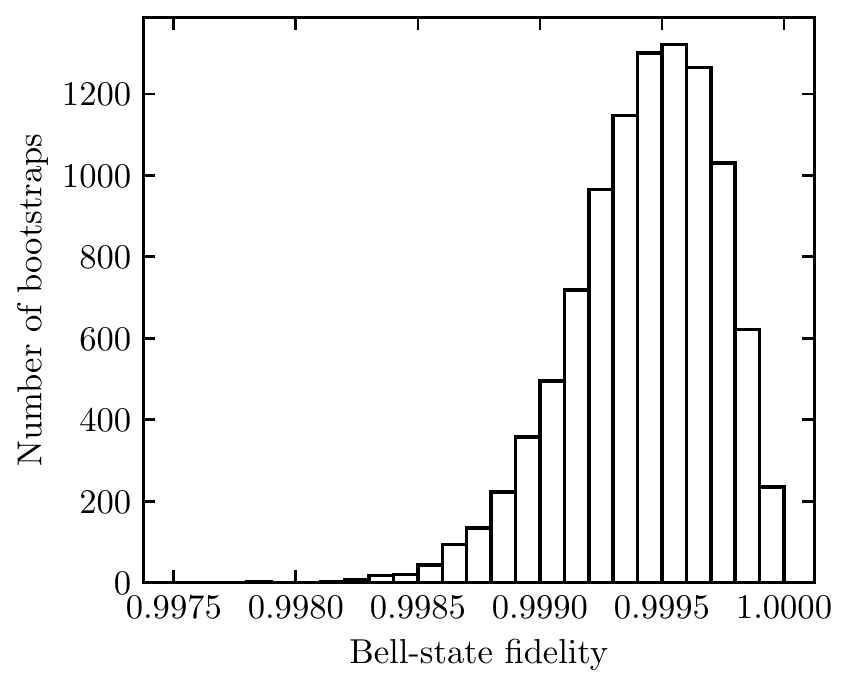}
	\caption{\label{fig:fidelity_histogram} Histogram of bootstrapped Bell-state fidelities.}
\end{figure}
Figure~\ref{fig:fidelity_histogram} gives a histogram of the Bell-state fidelities obtained from the bootstrapping procedure defined in the main text.

\subsection{Gate Errors}
1-qubit rotation fidelity is limited by qubit frequency noise, laser phase noise, and laser intensity fluctuations. Qubit frequency noise results from variations in the background magnetic field; our qubit is first-order sensitive to these variations which arise in our lab from changes such as the movement of a chair or the motion of a nearby elevator. Although the incorporation of a spin echo into our gate should alleviate the influence of these frequency variations, 1-qubit rotations (of finite duration) still suffer to some extent. To further reduce these errors, we interleave calibrations of the qubit frequency within our gate experiments, performing a Ramsey frequency calibration experiment between each repetition of the gate experiment. In this way the frequency is stabilized to approximately $\pm$250~Hz, a level at which it is expected to contribute an error of only $4\times 10^{-6}$ to the 1-qubit pulses. We use a NdFeB permanent magnet to minimize fast fluctuations of the $1.07~\text{mT}$ experiment bias field at the expense of slow (correctable) thermal drifts. The remaining error is likely caused by faster phase noise, intensity noise, or fast ambient magnetic field fluctuations.

Additional errors arise due to variations in the phase of the 729 nm laser between the various 1-qubit pulses. Such phase fluctuations can occur due to instability in the laser itself, vibrations of the mirrors along the beampath, and motion of the ion or vacuum chamber. Ramsey coherence measurements performed by repeating the gate experiment [pulses (1), (3), and (5) of Fig.~\ref{fig:pulse_profiles}(b)] while extinguishing the ODF beams indicate that this error contribution is $<2\times 10^{-4}$. Here a marked improvement was observed after the installation of an electro-optic modulator in the output path of the 729 nm laser for phase-noise stabilization.

Fundamentally the fidelity is limited by metastable decay from the $D_{5/2}$ level as well as by spontaneous photon scattering from the ODF beams. These contributions are discussed in detail in Ref.~\cite{sawyer_2021}. We estimate a photon scattering error of $1.1\times 10^{-5}$ at our c.m. frequency of 2.5~MHz. We estimate the metastable decay error as follows. Assuming on average that 50\% of the population lies in the $D_{5/2}$ level until the detection laser is turned on 50~$\mu$s later (this includes the time required for the parity analysis pulse as well as some additional short delays), there is a metastable decay error of $2\times 0.5\times 50\:\mu\mathrm{s}/1.174\:\mathrm{s}=4.3\times 10^{-5}$. In addition, the ions are illuminated with the detection laser for 50~$\mu$s before we begin counting photons (this allows the intensity of the laser to stabilize). However, due to the Zeno effect, the metastable decay rate during this second interval is roughly half its unperturbed value, leading to an error of $2.1\times 10^{-5}$ (we have confirmed this reduction in decay rate via a full simulation of the Schr{\"o}dinger equation including all the relevant levels in $^{40}$Ca$^{+}$). Therefore, the total metastable decay error is approximately $6\times 10^{-5}$. A closely related source of error is metastable decay during the measurement itself. In the case of 100~$\mu$s (200~$\mu$s) measurement duration used for the data in Fig.~\ref{fig:parity_amplitude} (Fig.~\ref{fig:population_scan}), we calculate a lifetime-limited detection error of $4.3\times 10^{-5}$ ($8.6\times 10^{-5}$), again considering the reduced decay rate during illumination. These are likely overestimates, because decays that occur later during the measurement interval should only minimally impact the photon count histograms. We have chosen not to SPAM-correct our reported fidelity in order to reduce statistical complexity.

To initialize to $\qubitdowndown$, we employ optical pumping with a circularly polarized 397 nm laser beam followed by seven cycles of two-step optical pumping with the 729~nm and 854~nm laser beams exciting along $S_{1/2}$~$(m_j=-1/2)\rightarrow D_{5/2}$~$(m_j= 1/2)\rightarrow P_{3/2}$. The initial 397~nm pumping step yields an error of $\sim1\times10^{-3}$, and the following steps further reduce this error to $<10^{-6}$, which we exclude from the error budget table in the Letter.

In the Lamb-Dicke limit there is no dependence of the gate on ion temperature~\cite{sorensen_2000}. A derivation of the higher-order deviations from perfect Lamb-Dicke assumptions reveals a gate error of $\epsilon_g=(\pi^2/4) \eta^4 \bar{n}(2\bar{n}+1)$, where $\bar{n}$ represents the thermal mean number of quanta in and $\eta$ is the Lamb-Dicke parameter of the mode \cite{sorensen_2000,ballance_2016}. This dependence, which applies independently for both the BM and c.m. finite-temperature contributions, is confirmed via numerical simulations. We estimate $\bar{n}_\mathrm{BM}<0.1$, $\bar{n}_\mathrm{c.m.}<0.1$, $\eta_\mathrm{BM}=0.063$, $\eta_\mathrm{c.m.}=0.083$ so that $\epsilon_g<2\times 10^{-5}$.

Heating of the gate mode during operation is unavoidable. We operate the gate near the BM mode instead of near the c.m. mode because of the former's significantly lower $<1.4$~quanta/s heating rate (the c.m. heating rate is $33(14)$~quanta/s). This translates into a gate error of $\epsilon_g=\dot{\bar{n}} \tau_g/(2 K)=7\times 10^{-6}$ ($\tau_g$ here represents the sum of the two ODF pulse durations, and $K=2$ is the number of loops in phase space) \cite{sorensen_2000}.

Variations in the trapping potential can lead to errors via two possible mechanisms. Fast electric field variations can change the position of the trap potential within the Ramsey sequence [Fig~\ref{fig:pulse_profiles}(b)]. This manifests as phase noise both on the optical-dipole force interaction and on the 1-qubit rotations \cite{pino_2020}. For the 2~Hz low-pass filters in use on our dc electrodes, we expect this error contribution is negligible. However, it may become relevant if filters with a higher cutoff frequency are used in the future. 

Slower variations in the trapping potential (for example due to thermal drift in the DAC electronics or due to slow charging or discharging of the trap surfaces) can cause variations in the trap frequency from shot to shot of the Ramsey sequence. We have characterized the trap frequency by driving ion motion with a radio-frequency electric field and detecting the ions’ response. For an 8~ms pulse, the resulting frequency response is well fit to a Gaussian with a standard deviation of $\Delta \omega/2\pi=63(11)$~Hz. This predicts an error $\epsilon_g=(\pi^2/4) (\Delta \omega/\delta)^2\approx1\times 10^{-6}$ for our gate detuning $\delta/2\pi=114$~kHz. It is likely that this frequency uncertainty is dominated by Kerr cross-coupling between the (Doppler-cooled) radial rocking modes and the BM mode \cite{nie_2009}.%[S1].

To lowest order, fluctuations in laser intensity between the two ODF pulses lead to errors because of the associated uncompensated phase accumulation. To estimate the effect of gate laser beam power fluctuations, we measure the power stability of the two 532~nm laser beams separately with the photodetectors used for power stabilization. From this, we calculate an Allan variance of power fluctuations at $\tau_\mathrm{ODF}=12$~$\mu$s that is consistent with the photon shot noise at each detector. The resulting error contribution is bounded from above by $1\times10^{-5}$~\cite{sawyer_2021}, which is included with the first entry in Table~\ref{table:error_budget}.  

In addition to power instability, beam pointing or ion position variations will also cause intensity to fluctuate. We measure this experimentally by performing pulses (1)-(5) while illuminating the ions with only one of the two ODF beams. Performing these measurements with sufficiently many repetitions to accumulate good statistics is time consuming, but we routinely observe no errors in $1\,000$ repetitions, bounding this error source to $<10^{-3}$. Extending this characterization to include a similar number of experiments ($\sim10\,000$) as are used for Bell-state verification would provide a more restrictive bound. Slow (shot-to-shot) variations in ODF laser-beam intensity will also contribute to the Bell-state infidelity. This error source is difficult to quantify independently of the entangling gate but is reflected in the first entry of Table~\ref{table:error_budget}. 

\end{document}